\documentclass[prb,aps,twocolumn, superscriptaddress, longbibliography,floatfix]{revtex4-2}
\usepackage[
  colorlinks=true,
  urlcolor=blue,
  linkcolor=blue,
  citecolor=blue,
  bookmarks=false,
  pdftitle={},
  pdfauthor={}
]{hyperref}
\usepackage{import}
\usepackage{subfiles}
\usepackage{graphicx}
\usepackage{bm}
\usepackage{amsmath}
\usepackage{amssymb}
\usepackage{mathtools}
\usepackage{amsfonts}
\usepackage[all]{xy}
\usepackage{xspace}
\usepackage{accents}
\usepackage{stmaryrd}
\usepackage{tabularx}
\usepackage{extarrows}
\usepackage{float}
\usepackage{multirow}
\usepackage{makecell}
\usepackage[capitalise]{cleveref}
\usepackage[dvipsnames]{xcolor}
\usepackage[normalem]{ulem}
\usepackage{soul}
\usepackage{pdfpages}
\usepackage{cancel}
\makeatletter
\AtBeginDocument{\let\LS@rot\@undefined}
\makeatother

\usepackage{changes}
\definechangesauthor[color=Red]{SR}

\begin{document}

\title{Universal magnetotunnel conductance at a Weyl semimetal–layered\\ Chern insulator junction}
\author{Nirnoy Basak}
\affiliation{Department of Physics, Indian Institute of Technology Kanpur, Kalyanpur, Uttar Pradesh 208016, India}
\author{Sumathi Rao}
\affiliation{International Centre for Theoretical Sciences, Tata Institute of Fundamental Research, Bengaluru 560089, India}
\author{Faruk Abdulla}
\affiliation{Physics Department, Technion - Israel Institute of Technology, Haifa 32000, Israel}
\affiliation{The Helen Diller Quantum Center, Technion, Haifa 32000, Israel}

\begin{abstract}
We investigate electronic transport across a junction between a Weyl semimetal 
(WSM) and a layered Chern insulator (LCI) in the presence of a magnetic field perpendicular 
to the interface. The topological mismatch between the gapless Weyl 
semimetal and the momentum-resolved chiral edge modes of the layered Chern 
insulator leads to interface  Fermi-arc states with a qualitatively distinct 
connectivity: unlike WSM–WSM junctions, the interface Fermi arcs are forced 
to reconnect through the Brillouin-zone boundary rather than terminating at the
projections of the  Weyl nodes. We analyze the spectrum and compute the magneto tunnel conductance mediated by the interface-localized states. We find that the 
conductance increases linearly with magnetic field at low fields and saturates beyond a 
critical field to a constant value that is independent 
of microscopic details such as interface coupling, arc geometry, and 
lattice-scale parameters. This universal saturation reflects a transport 
mechanism governed by the topological charge pumping associated with the Chern 
layers, rather than magnetic breakdown between Fermi arcs. We further show that, 
under specific conditions, a junction between two distinct Weyl semimetals can 
exhibit a similar saturation behavior, thereby clarifying the topological 
origin of the observed universality. 

\end{abstract}

\maketitle

\section{Introduction}

Weyl semimetals (WSMs) host topologically protected band-touching points 
in the three-dimensional Brillouin zone, which act as monopoles of Berry 
curvature and give rise to a variety of unconventional electronic phenomena, 
including the chiral anomaly and anomalous magnetotransport responses 
\cite{Murakami_2007, Wan_Savrasov_2011, Burkov_Balents_2011, Lv_Ding_2015a,
Lv_Ding_2015b, Xu_Hasan_2015a, Xu_Hasan_2015b,  Aji_2012, 
Zyuzin_Burkov_2012, Son_Spivak_2013, Gorbar_Miransky_2014, Burkov_2015}. 
A defining feature of WSMs is the presence of Fermi-arc surface states—open 
segments in the surface Brillouin zone that connect the projections of bulk 
Weyl nodes of opposite chirality\cite{Burkov_Balents_2011, Xu_Hasan_2015a, Xu_Hasan_2015b, 
Wan_Savrasov_2011}. These states represent a striking 
manifestation of bulk–boundary correspondence and play a central role in 
transport phenomena unique to Weyl systems.

Interfaces between two Weyl semimetals provide a particularly rich setting 
for exploring the interplay between bulk topology, surface states, and transport. 
When the projections of the Weyl nodes on the two sides of the interface do 
not coincide, the interface generically hosts Fermi-arc states formed from 
the hybridization of surface arcs of the individual WSMs \cite{Dwivedi_2018, 
Ishida_2018, Murthy_2020, faruk2021farecon, Francesco_2022, Mathur2023, 
Ritajit2023}. When a magnetic field is applied perpendicular to the interface, 
the interface Fermi arcs enable transport across the junction, with the current 
carried by the chiral zeroth Landau levels \cite{chau2023magnetic} being 
transmitted between the two WSMs. Depending on how the 
arcs connect the projected Weyl nodes, the resulting transport can exhibit 
perfect transmission or perfect reflection at low fields, followed by a 
saturation of the magnetotunnel conductance at high fields due to magnetic 
breakdown at close encounters of Fermi arcs \cite{chau2023magnetic, chaou2023quantum, 
Nirnoy_2024, Maxim_2026}. In such WSM–WSM junctions, 
the value of the saturated conductance is controlled by microscopic details of 
the interface Fermi-arc geometry, such as the arc separation, intersection 
angle, and hybridization gap.

More recently, it has been shown that multiple close encounters between 
interface Fermi arcs can lead to non–Shubnikov–de Haas quantum oscillations 
in the magnetotunnel conductance, originating from momentum-space Aharonov–Bohm 
interference of semiclassical trajectories along the arcs \cite{chaou2023quantum}. 
These studies 
have firmly established interface Fermi arcs as active carriers of 
magnetotransport and have clarified the role of magnetic breakdown in 
determining the high-field response of WSM–WSM junctions.

\begin{figure}[t]
\centering
\includegraphics[width=0.9\columnwidth]{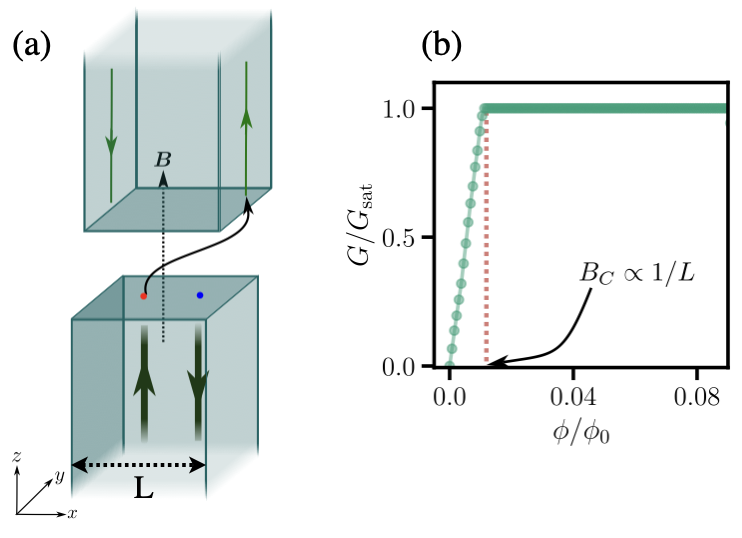}
\caption{(a) \textbf{Schematic diagram of the set up:} A junction of a 
WSM (bottom) and a LCI slab (top) is placed in a perpendicular magnetic field. 
In the WSM slab, currents carried by only the chiral Landau level states 
(green arrows) can transmits to LCI slab. The interface states 
(see Fig. \ref{fig:FA_LCI_WSM}) 
which are localized along $z$ but are propagating in the plane of the interface 
redirects the current carried by the chiral LL states to the chiral edge states 
(green arrows on the $y$-$z$ surfaces) of the LCI slab. (b) \textbf{Summary of 
the main result:} There are two regimes for magneto-tunnel conductance $G$- a 
linear regime which is followed by saturation to a universal value $G_{sat} 
= \frac{e^2}{h}L_y$ which is independent of any microscopic details of the system. 
The tunnel conductance in the linear regime $G = \frac{e^3}{h^2} {\bf A} 
\cdot {\bf B}$, where $A$ is the area of interface, is also independent of any 
microscopic details of the system. }
\label{Fig:Setup}
\end{figure}

In this work, we explore a qualitatively different interface geometry: a 
junction between a WSM and a layered Chern insulator (LCI). The 
setup is demonstrated in Figure \ref{Fig:Setup}.
An LCI may be viewed as a stack of two-dimensional Chern 
insulators, characterized by a nonzero momentum-resolved Chern number and 
supporting chiral edge modes for each conserved momentum along the stacking 
direction. Unlike a Weyl semimetal, the LCI is fully gapped in the bulk 
and does not host Weyl nodes. The junction between a WSM and an LCI therefore 
represents a fundamentally distinct topological interface, where gapless 
Weyl fermions on one side couple to momentum-resolved chiral edge states 
on the other.

We show that this topological mismatch leads to interface Fermi-arc states 
with a connectivity that has no analog in WSM–WSM junctions. In particular, 
the interface Fermi arcs are not terminated by Weyl-node projections on both 
ends. Instead, they are forced to reconnect through the boundary of the 
Brillouin zone, reflecting the presence of a nontrivial Chern number on the 
LCI side. This connectivity is topologically enforced and persists independently 
of microscopic details such as interface coupling strength or lattice-scale parameters.

We investigate the consequences of this unconventional interface spectrum 
for magnetotransport across the junction. Focusing on a magnetic field 
perpendicular to the interface, we compute the magnetotunnel conductance 
mediated by the interface-localized states. We find that, while the conductance 
increases linearly with magnetic field at low fields, it saturates beyond 
a critical field $B_c$ to a constant value that is independent of microscopic
details of the system, including the precise shape and length of the interface 
Fermi arcs (See Fig.~\ref{Fig:Setup} for the setup and the main result). 
We find that the critical field at which saturation occurs scales as 
$B_c \propto 1/L$. Thus, in the thermodynamic limit, the tunnel conductance 
saturates to a universal value even for a very small field. 
This behavior stands in sharp contrast to WSM–WSM junctions, where conductance 
saturation arises from magnetic breakdown and depends explicitly on interface 
geometry. In the WSM–LCI junction, the universal saturation instead reflects 
a transport mechanism governed by topological charge pumping associated with 
the Chern layers of the LCI.

Finally, we show that a similar saturation behavior can also arise in 
junctions between two distinct WSMs, provided specific conditions 
are met that effectively mimic the role of a momentum-resolved Chern insulator. 
This comparison clarifies the topological origin of the universality observed 
in the WSM–LCI junction and highlights the essential role played by the Chern 
character of the interface states.

The remainder of this paper is organized as follows. In Sec. II, we introduce 
the model for the WSM–LCI junction and analyze the interface-localized states. 
In Sec. III, we compute the magnetotunnel conductance and demonstrate the 
emergence of universal saturation at high magnetic fields. In Sec. IV, we 
discuss the relation to WSM–WSM junctions and identify the conditions under 
which similar behavior can arise. We conclude in Sec. V with a summary and outlook.

\section{Interface states for WSM-LCI junction.}

\subsection{Model}

We begin from a minimal two-band lattice model on a cubic lattice that 
realizes a Weyl semimetal phase with two Weyl nodes separated along the 
$k_x$ direction. The Bloch Hamiltonian is written as
\begin{align}
    H_{\text{WSM}}({\bf k}) = M({\bf k}) \sigma_x +  v_y\sin{k_ya}\sigma_y + v_z\sin{k_za}\sigma_z
\end{align}
with $M(k) = \frac{v_x}{\sin{k_0a}}(2 + m - \cos{k_xa} - \cos{k_ya} - \cos{k_za})$, where $\sigma_{x,y,z}$ are Pauli matrices acting on the pseudospin space, and $a$ is 
the lattice constant which we set to unity. For an 
appropriate range of the parameter $m$, the bulk spectrum hosts two Weyl  nodes 
located at ${\bf k} = (\pm k_0, 0, 0)$, with $k_0 = \cos^{-1}(m)$. In what follows, 
we set $k_0=\pi/2$ so that the two Weyl nodes are at maximum separation. 
These nodes carry opposite chirality and give rise to Fermi-arc surface states on 
boundaries perpendicular to the $z$ and $y$ directions. In $H_{\text{WSM}}$, the 
$v_i$ ($i=x, y,z$) are  the velocity components of the Weyl fermion along 
the $i$'th direction and here we define a ratio $v_r=v_y/v_x$ which plays  
an important role in transport and will be discussed later.

The LCI  phase is obtained from the same parent Hamiltonian by continuously 
tuning parameters such that the two Weyl nodes move towards the boundary of 
the Brillouin zone along $k_x$ and annihilate pairwise. After annihilation, 
the bulk spectrum becomes fully gapped. However, for each fixed value of 
momentum along the $k_x$ direction, the effective two-dimensional 
subsystem carries a nonzero Chern number. The resulting phase can therefore 
be viewed as a stack of two-dimensional Chern insulators and supports 
chiral boundary modes for each momentum in the topological regime. In 
this sense, the LCI is the gapped descendant of the Weyl semimetal 
obtained via Weyl node annihilation at the BZ boundary.

To form the junction, we consider a slab geometry along the $z$-direction 
and place the interface at $z=0$. The WSM occupies $z<0$, while the layered 
Chern insulator occupies $z>0$. The translation symmetry along the $x$ and $y$ 
directions is preserved and ($k_x,k_y$) remain good quantum numbers. The interface 
 Brillouin zone (IBZ)  is, therefore, spanned by $(k_x,k_y)$.

Prior to forming the junction, we rotate the LCI slab by $90^\circ$ about 
the $z$-axis. This choice serves two purposes. First, it ensures that the 
underlying cubic lattices of the two slabs remain aligned at the interface, 
so that no additional superlattice structure is required. For a generic 
rotation angle, the two lattices would become incommensurate, 
necessitating the construction of a supercell to define a common interface 
Brillouin zone. Such a superlattice treatment would substantially increase 
the computational cost without introducing qualitatively new physics. Second, 
the $90^\circ$ rotation ensures that, at zero interface coupling, the Fermi 
arcs of the WSM and the chiral edge modes of the LCI intersect at right angles 
in the IBZ. This orthogonal crossing provides a transparent starting point 
for analyzing the reconstruction of interface states upon hybridization. 
The $90^\circ$ rotation also ensures that the transverse surface states 
which are on ${\hat x}$-${\hat z}$ plane on the WSM are on the 
${\hat y}$-${\hat z}$ plane  on the LCI and are not connected.  

\begin{figure}
 \centering
 \includegraphics[width=1\linewidth]{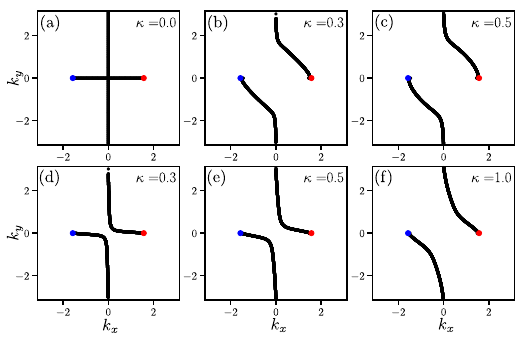}
 \caption{Shows interface localized states of the WSM-LCI junction for 
 a few values of tunneling coupling $\kappa$. (a) When $\kappa=0$, 
 the interface hosts the Fermi arc surface states from the WSM side 
 and the chiral egde states from the LCI side. At finite coupling, they hybridize. 
 The interface-localized states form continuous open contour that connects 
 the projected Weyl nodes on the WSM side through the boundary of the the 
 interface BZ. The first and second row are for two different values of 
 the bulk parameter $v_r=0.2$ and 1 respectively. } 
 \label{fig:FA_LCI_WSM}
\end{figure}

\subsection{Interface Coupling}

The full Hamiltonian of junction is 
\begin{align}
    H = H_{\text{WSM}} + H_{\text{LCI}} + H_{\text{int}}
\end{align}
where $H_{\text{WSM}}$ and $H_{\text{LCI}}$ describe the two 
subsystems in slab geometry along $z$. The interface coupling 
connects the top layer of the WSM to the bottom layer of the LCI 
and is parametrized by two real parameters $\kappa$ and $u$
\begin{align}
   H_{\text{int}} = \kappa \sum_{{\bf k}_{\parallel}} \left( c^{\dagger}_{\text{WSM}, z=0} 
   ({\bf k}_{\parallel})(u\sigma_x + i\sigma_z)  c_{\text{LCI}, z=0}({\bf k}_{\parallel}) + H.c \right)
\end{align}
with ${\bf k}_{\parallel}=(k_x, k_y)$. In general the coupling matrix, 
can also include a term proportional to $\sigma_y$. As previously discussed in 
Ref.~\onlinecite{faruk2021farecon}, such a term would not have effect on the Fermi 
arc reconstruction 
at the interface because the surface states localized at the interface of the top/bottom 
slabs are up/down pseudospin polarized along the $y$ direction in the Pauli space
and hence a $\sigma_y$ term cannot contribute in hybridization. 
The parameter $u$ indicates an 
imbalance between pseudospin-preserving and pseudospin-flipping hoppings 
across the interface in contrast to the bulk hopping. In what follows 
we will fix $u=0.5$ and use $\kappa$ as a control parameter which measures
the strength of hybridization across the interface. In the decoupled limit 
$\kappa=0$, the interface localized states consist of the Fermi arc states of 
the WSM and the chiral edge states of LCI slabs.  
For finite $\kappa$, the interface states are obtained from exact 
diagonalization of the full coupled Hamiltonian for each $(k_x,k_y)$.

\subsection{Interface localized states}

To determine the interface spectrum, we numerically diagonalize the slab 
Hamiltonian at fixed $(k_x,k_y)$ and identify states localized near $z=0$ 
via their spatial probability distribution. The resulting dispersion 
$E(k_x,k_y)$ reveals interface-localized bands lying within the bulk 
gap of the LCI and within the projected Weyl node region of the WSM.

At $\kappa=0$, the IBZ contains two distinct sets of gapless states: 
(i) Fermi arcs from the WSM surface connecting the projections of the 
Weyl nodes separated along $k_x$, and (ii) chiral edge modes from the 
LCI which disperse along $k_x$. These states cross orthogonally without 
hybridization.

Upon turning on a finite interface coupling $\kappa$, hybridization at 
the crossings reconstructs the spectrum. We find that the resulting 
interface-localized states form continuous open contours that connect 
the projected Weyl nodes on the WSM side through the boundary of the 
BZ (see Fig. \ref{fig:FA_LCI_WSM}). Because the LCI does not host Weyl 
nodes, the interface Fermi arcs cannot terminate on node projections on both ends. 
Instead, they are forced to reconnect via the IBZ boundary, reflecting 
the nontrivial Chern number inherited from the Weyl-node annihilation 
process.

This connectivity is robust against variations of $\kappa$ and other 
microscopic details, provided the bulk topology of the two subsystems 
remains unchanged. As we show in the following section, this distinctive
Fermi-arc structure plays a central role in determining the magnetotransport 
properties of the WSM–LCI junction.

\begin{figure}[tbph]
 \centering
 \includegraphics[width=1.0\linewidth]{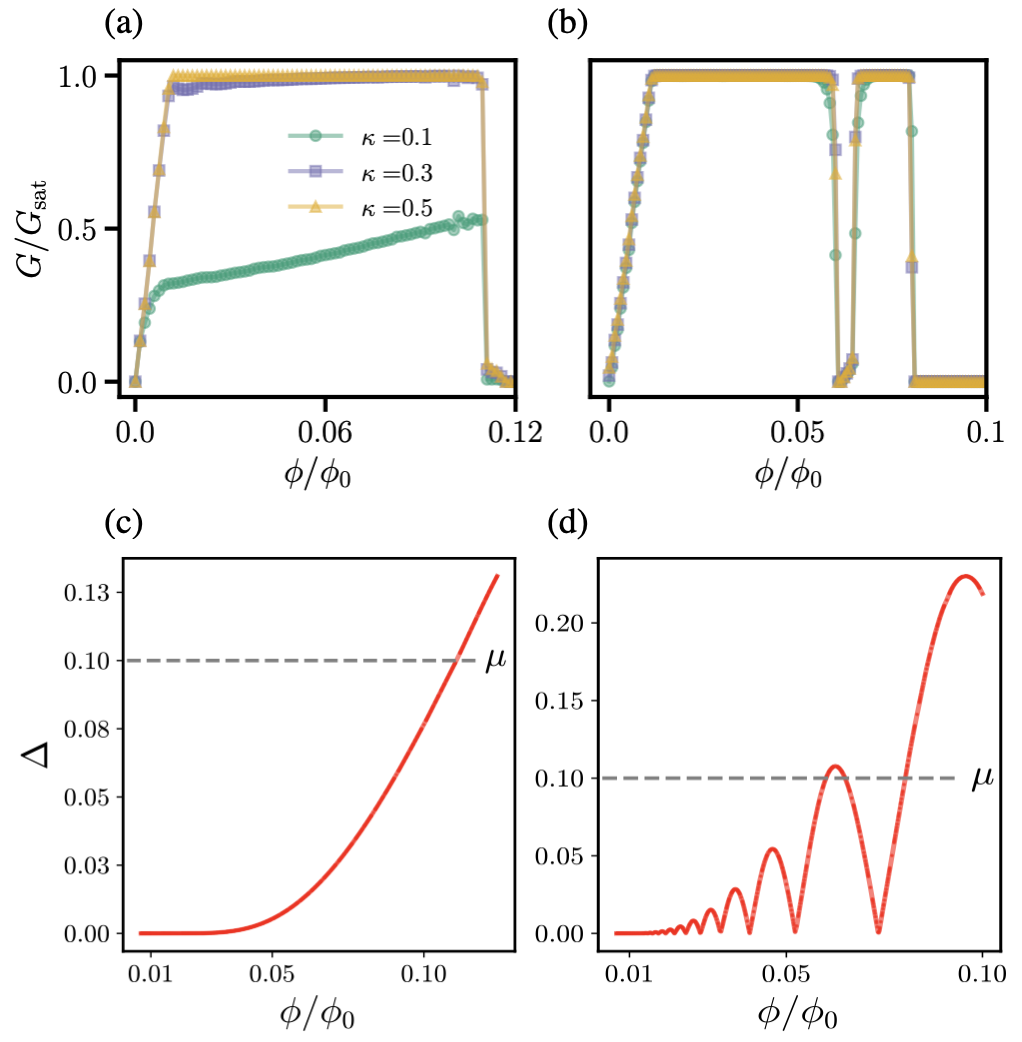}
 \caption{(a) and (b) show magneto-tunnel conductance across the WSM-LCI junction
 for several values of interface coupling $\kappa$ as a function of flux through the 
 interface. In (a) the bulk parameter $v_r=v_y/v_x$=1.0 and in (b) it is 0.2. 
 The system size along the x and y direction are $L_{x}=L_y=100$. The chemical 
 potential (measured relative to the energy of the Weyl point) is $\mu=0.1$. 
 Unless interface coupling parameter $\kappa$ is small (see main text for 
 an explanation), the conductance saturates 
 to a universal value $G_{sat}=\frac{e^2}{h} L_y$ at flux $\phi/\phi_0 = 1/L_x$. 
 (c) and (d) show the bulk gap $\Delta$ in the WSM induced by the magnetic field for 
 $v_r=1.0$ and 0.2 respectively. Whenever the field induced bulk gap $\Delta$ is 
 larger than the chemical potential, the tunnel conductance drops to zero. } 
 \label{Fig:Cond_WSM_LCI}
\end{figure}

\section{Magneto-Tunnel Conductance across junction }
\label{Sec:TCinAofMF}

\subsection{Zero-field transport}
We first discuss transport in the absence of a magnetic field. Since the 
LCI is fully gapped in the bulk, only its surface states can participate 
in low-energy transport. On the WSM side, surface Fermi arcs exist on 
boundaries ($y=0, L_y$) perpendicular to the z-direction. In the slab 
geometry considered here, current injected from the WSM side is carried 
by these surface states.

At zero field, bulk states in the WSM do not contribute to transmission 
across the junction at low energy, as there is no direct bulk channel in 
the LCI to receive them. Instead, transport proceeds entirely through 
interface-localized states that are exponentially localized in the 
$z$-direction but propagate in the $(x,y)$-plane. As discussed in our 
previous work in Ref. \cite{Nirnoy_2024}, the current 
arriving at the interface from the WSM surface is redirected along 
these interface states and subsequently transferred to the chiral 
edge modes of the LCI slab. 

Thus, at $B=0$, the conductance across the junction is determined solely 
by the overlap and hybridization between surface Fermi arcs of the WSM and 
chiral edge modes of the LCI. The bulk topology of the WSM does not 
directly contribute to transport in this limit.



\subsection{Transport in perpendicular magnetic field}

The situation changes dramatically in the presence of a magnetic field 
${\bf B}$  parallel to  $\hat{z}$ and  perpendicular to the interface. The field quantizes 
 the bulk states of the WSM to form Landau levels. In particular, each Weyl node 
generates a chiral zeroth Landau level (LL) that disperses linearly along 
the field direction. These chiral LLs provide bulk channels that carry 
current toward the interface.

In contrast, the LCI remains fully gapped in the bulk even in the presence 
of the magnetic field. Its low-energy transport continues to be carried
by chiral edge modes associated with the momentum-resolved Chern number.

The essential mechanism for magnetotransport across the junction is therefore 
the following: current carried by the chiral Landau level in the WSM is 
redirected at the interface into chiral edge modes of the LCI. As discussed in
previous works \cite{chaou2023quantum, chau2023magnetic, Nirnoy_2024}, this 
redirection occurs via the interface-localized 
states identified in Sec. II. The interface states act as a momentum-space 
bridge between the bulk chiral LL of the WSM and the boundary modes of the LCI.


In the presence of the magnetic field, scattering processes at the 
interface become possible. In particular, an electron propagating along one 
Fermi arc may tunnel to another arc and be reflected back into the WSM. 
Such processes correspond to magnetic breakdown between interface states
and  suppress the conductance by an exponentially small factor
$(1-e^{-B_0/B})$, where $B_0 \propto q^2_{FAs}$ and  $q_{FAs}$ 
denotes the momentum space distance between the two Fermi arcs at the interface,
as shown earlier in Refs.~ \onlinecite{chau2023magnetic, chaou2023quantum}. 

As shown below, there exists a characteristic field $B_c$, determined 
solely by the system size perpendicular to the transport direction. When 
$B_c \lesssim  B_0$, the effect of such scattering processes becomes negligible, 
and the tunnel conductance saturates to a constant value independent 
of microscopic details such as the interface coupling strength $\kappa$, 
the precise geometry of the Fermi arcs, or lattice-scale band parameters.

\subsection{Field dependence and saturation of conductance}

The magnetic field introduces a large degeneracy of chiral LL modes. 
For a system of dimensions $L_x \times L_y$, the number of chiral 
LL channels in the WSM is
\begin{align}
N_{\text{LL}} = B L_x L_y/\phi_0,
\end{align}
reflecting the Landau degeneracy proportional to the magnetic flux 
(in units of flux quantum $\phi_0=h/e$) through the interface plane.

On the LCI side, however, the number of available chiral edge modes 
is determined solely by the Chern character of the layered insulator 
and by the system size perpendicular to the direction of propagation. 
In the present geometry, the number of LCI chiral edge modes  
$N_{\text{edge}} = L_y$ (here, the lattice constant has been set to unity), and is 
independent of the magnetic field.

This mismatch in channel scaling plays a central role in determining the 
magnetotunnel conductance. For weak magnetic fields, the number of incoming 
chiral LL modes is smaller than or comparable to the number of available 
LCI edge channels. In this regime, the conductance increases approximately 
linearly with B, reflecting the increasing number of transmitting chiral 
LL modes.

\begin{figure}
 \centering
 \includegraphics[width=0.8\linewidth]{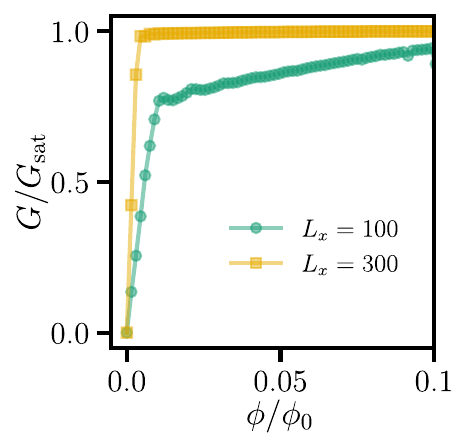}
 \caption{Shows the system-size dependence of magneto-tunnel conductance 
 across the WSM-LCI junction. For small tunnel coupling $\kappa=0.2$, the 
 scattering between the Fermi arcs at the interface is significant and hence
 the conductance does not saturate to universal value $G_{\mathrm{sat}}
 =\frac{e^2}{h}L_{y}$ for small system size along $x$ direction. However for 
 larger system size $L_x=300$, the conductance saturates to the universal value
 $G_{sat}$ at a flux value $\phi/\phi_{0}=1/L_{x}$. The system size along 
 the $y$ direction is fixed to $L_y=50$. } 
 \label{Fig:Cond_LCI_WSM_L}
\end{figure}

As the magnetic field is increased further, the number of WSM chiral LL modes 
eventually exceeds the number of available LCI edge modes. Beyond this point, 
additional LL channels cannot be accommodated by independent outgoing edge 
modes on the LCI side, and hence the transport becomes limited by the number 
of available LCI edge channels. The conductance therefore saturates to a 
constant value determined solely by the number of LCI chiral edge modes 
\begin{align}
    G_{sat} = \frac{e^2}{h} L_y,  
\end{align}
and becomes independent  of further increases in the magnetic field. 
This behavior of the tunnel conductance as a function of magnetic field 
(computed by employing KWANT simmulation \cite{groth2023kwant}) is 
demonstrated in Fig. \ref{Fig:Cond_WSM_LCI}. 

The crossover field $B_c$ at which saturation sets in is obtained by 
equating the number of chiral LL modes to the number of edge channels
$B_cL_x L_y/\phi_0 = L_y$, resulting in $ B_c = \phi_0/L_x$. Thus, the onset 
of saturation shifts to lower magnetic fields for wider systems. 
Importantly, the maximum  conductance is independent of microscopic 
details such as the interface coupling parameter $\kappa$, the precise
geometry of the Fermi arcs, or lattice-scale band parameters. Once the 
magnetic field is sufficiently strong, transport is controlled entirely 
by the topological constraint imposed by the finite number of LCI chiral 
edge modes. The universality of the saturation therefore reflects the 
Chern character of the LCI rather than details of interface scattering.

We note that the conductance for small interface coupling, as shown in  Fig. \ref{Fig:Cond_WSM_LCI}(b), does  not saturate to the constant 
value $G_{sat}$. This is because the momentum-space distance 
$q_{FA}$ between the two Fermi arcs at the interface is small, 
 due to the stronger scattering when $B_0$ is small compared to $B_c$. 
Nevertheless, for any thermodynamically large system, the saturation 
field $B_c \to 0$ (due to the large  degeneracy of the chiral LLs), and 
hence the tunnel conductance should  saturate to $G_{sat}$ irrespective 
of the strength of the interface coupling parameter. This is demonstrated 
in Fig. \ref{Fig:Cond_LCI_WSM_L} by plotting the tunnel conductance for 
larger system sizes. 

We emphasize that Fermi arc states of the WSM, propagating towards the 
interface, also contribute to transport. However, in the tunneling 
conductance shown in Fig. \ref{Fig:Cond_WSM_LCI}, this contribution 
has been neglected. As demonstrated in our previous work \cite{Nirnoy_2024}, 
these states affect only the linear component of the tunneling conductance 
and does not modify the overall behavior discussed here. Additionally, for 
large system size in the thermodynamic limit, this contribution is negligible.


We notice that magneto tunnel conductance, shown in Fig. \ref{Fig:Cond_WSM_LCI}, 
behaves quite differently for the two values of $v_r=v_y/v_x=1.0$ and 0.2. 
This is somewhat surprising since the naive expectation would be for the tunnel 
conductance to  be  qualitatively similar when   
bulk parameters such as  the velocities $v_x$ and $v_y$  of the Weyl fermion are  
changed.  In particular, we note  that 
for $v_r=0.2<1$,  the conductance oscillates between saturation values and zero 
as a function of the magnetic field. This unusual behavior of the 
magneto-tunnel conductance can be seen to be  linked to subtle changes in 
the bulk band structure at low energies  of the WSM due to the magnetic field. 
Recent work have both theoretically 
established \cite{Chan_emergence_2017, Kim_breakdown_2017, Saykin_Landau2018,
Devakul_QO_2021, Abdulla_timerev_2022, Abdulla_pairwise_2024, Abdulla2026, 
Francesco_2024} as well as experimentally  observed \cite{Zhang_Jia_2017, Ramshaw_McDonald_2018} that an orbital magnetic 
field which is perpendicular  to the direction of separation of Weyl nodes 
of opposite chirality couple them and a finite gap $\Delta$ opens at the Weyl point. Consequently, the chiral LL becomes gapped at low energies. Therefore,  whenever  the chemical potential lies in the bulk gap, the conductance  drops to zero because 
there  are no states to carry the current. For $v_r=1.0\ge1$, the induced gap 
$\Delta$  monotonically increases with the field and for $v_r=0.2<1$ it oscillates
with field (see Figs. \ref{Fig:Cond_WSM_LCI}). Hence, by comparing 
\ref{Fig:Cond_WSM_LCI}(a) and (c), and (b) and (d), we see that  whenever the
chemical potential is below the gap, the conductance drops to zero.


\section{Universal conductance for WSM-WSM junction}

In the previous section we showed that the magnetotunnel conductance across 
a WSM–LCI junction saturates to a universal value at sufficiently large 
magnetic fields. In this section we demonstrate that a similar universal 
behavior can also arise in a junction formed between two WSMs under 
appropriate conditions.

As we discussed in the previous section, an external magnetic field which 
is applied along a direction perpendicular to the separation of a pair of 
Weyl nodes in momentum space can couple them and gap them pairwise. The resulting 
insulating state could be topologically nontrivial. As shown in recent 
works \cite{Abdulla_timerev_2022, Abdulla_pairwise_2024, Abdulla2026}, 
the  Landau quantization can lead to the formation  of either a normal 
insulating state or a layered Chern insulator depending 
on the separation between the Weyl nodes. For sufficiently large 
momentum-space separation between the nodes, the system effectively realizes 
a stack of two-dimensional Chern insulators. In this regime the 
magnetic field drives the system into a LCI phase where the Fermi arc 
states of the WSM evolve to the chiral edge states of LCI. 

\begin{figure}
 \centering
 \includegraphics[width=1\linewidth]{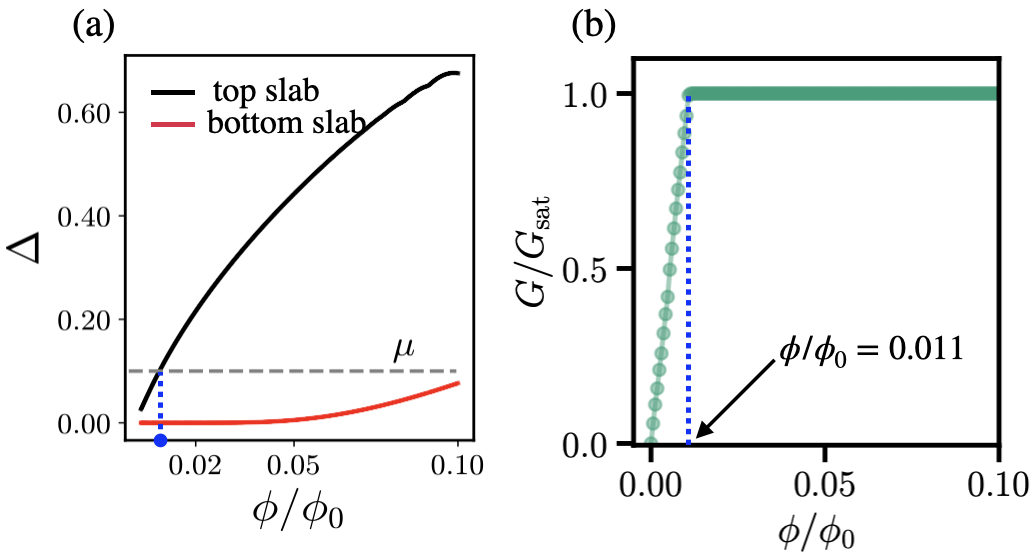}
 \caption{(a) Magnetic field induced gap in the bulk of the WSM slabs and (b) the 
 universal magneto-tunnel conductance across the junction. For the junction, 
 we adopt the model introduced in Eq. 1 of our previous work \cite{Nirnoy_2024} with 
 momentum space separation between the nodes $Q=2k_0=\pi$ and 5.6 for bottom and 
 top slabs respectively. We assumed tunnel coupling strength $\kappa=1.0$ and the 
 system size along $x$, $y$ directions to be $L_x=100$ and $L_y=50$ respectively. 
 For a chemical potential $\mu=0.1$, the junction effectively 
 behaves like a WSM-LCI junction for flux values beyond $\phi/\phi_0 \approx 0.011$ 
 (marked by a blue dot in (a)). (b) As expected, the conductance saturates to 
 universal value $G_{sat}$ at the corresponding flux value.   } 
 \label{Fig:Cond_WSM_WSM}
\end{figure}

Motivated by this observation, we consider a junction between two WSMs 
whose Weyl node separations are chosen differently. For the junction, 
we adopt the model introduced in Eq. 1 of our previous work \cite{Nirnoy_2024}, 
with the following modification: In the top WSM, 
the momentum-space separation between the Weyl nodes is chosen to be 
sufficiently large so that, in the presence of a perpendicular magnetic 
field, the system effectively behaves as a layered Chern insulator. 
In contrast, in the bottom WSM  is chosen to have a smaller node separation 
such that the magnetic-field–induced gap remains small and the chemical 
potential can be tuned so that the chiral zeroth Landau level persists 
in the bulk and continues to participate in transport

Under these conditions, once the magnetic field exceeds a certain 
threshold value, the junction effectively behaves as a WSM–LCI interface. 
The bulk states of the top WSM reorganize into momentum-resolved Chern 
insulating layers that support chiral edge modes, while the bottom WSM 
continues to host chiral Landau-level channels that carry current 
toward the interface. The transport mechanism therefore becomes 
identical to the one discussed in Sec. III: current carried by the 
chiral Landau level of the WSM is redirected at the interface into the 
chiral edge modes of the effective LCI phase.

As a consequence, the magnetotunnel conductance of the WSM–WSM 
junction exhibits the same qualitative behavior found for the WSM–LCI 
system (see Fig. \ref{Fig:Cond_WSM_WSM}). At low magnetic fields, the 
conductance increases with field 
due to the increasing degeneracy of the chiral Landau-level channels. 
Beyond a critical magnetic field $B_c$, however, the number of 
available outgoing chiral edge modes on the effective LCI side becomes 
the limiting factor for transport. The conductance therefore saturates 
to a constant value that is independent of further increases in magnetic field.


Importantly, the saturated conductance is insensitive to microscopic 
details such as the precise separation of Weyl nodes, the interface 
coupling strength, or other band-structure parameters. Instead, it is 
determined solely by the number of chiral edge modes supported by the 
effective layered Chern insulator. The emergence of this universal 
saturation thus reflects the topological nature of the interface transport.

This analysis shows that the universal conductance plateau identified 
in Sec. III is not restricted to junctions involving a genuine layered 
Chern insulator. Rather, it can also appear in appropriately tuned 
WSM–WSM junctions where one of the WSMs effectively realizes a layered 
Chern insulating phase in the presence of a magnetic field.

\section{Discussion and Conclusion}

In this work we investigated interface states and magnetotransport 
across a junction between a WSM and a LCI. Using a minimal two-band 
lattice model, we constructed the junction in a slab geometry and 
determined the interface-localized states via exact diagonalization. 
We found that the hybridization between the surface Fermi arcs of the WSM 
and the chiral edge modes of the LCI produces reconstructed interface 
states with a distinct connectivity in the interface Brillouin zone. 
In particular, since the LCI does not host Weyl nodes, the interface 
Fermi arcs cannot terminate on the node projection on the LCI side of the 
interface and instead reconnect through the boundary of the Brillouin 
zone, reflecting the Chern character of the LCI inherited from the 
Weyl-node annihilation process.

We then studied magnetotunnel transport across the junction in the 
presence of a magnetic field perpendicular to the interface. While 
transport at zero field is mediated only by surface states, a magnetic 
field introduces chiral  Landau-level channels in the bulk of the 
WSM that carry current toward the interface. These modes are 
redirected through the interface states into the chiral edge modes 
of the LCI. Because the number of incoming chiral Landau-level modes 
increases with magnetic field whereas the number of available edge 
channels in the LCI remain fixed, the magnetoconductance increases 
at low fields but eventually saturates to a constant value. The 
crossover field scales as $B_c \propto 1/L$, and the saturated 
conductance depends only on the number of LCI edge modes, making 
it independent of microscopic details of the interface.

We further showed that similar universal conductance saturation 
can also arise in junctions between two WSMs when a 
magnetic field effectively drives one of them into a LCI regime. 

Here we briefly comment on the robustness of the Fermi-arc–mediated 
magnetotransport across the interface in the presence of disorder. 
Interface disorder introduces a finite lifetime $\tau_{\mathrm{life}}$, 
which affects transport by enabling scattering between Fermi-arc states. 
As shown recently in Ref.~\onlinecite{Maxim_2026}, this leads to the 
emergence of a characteristic field scale $B_0^*$, defined by the 
condition that the Fermi-arc dwell time becomes comparable to $\tau_{\mathrm{life}}$.

In the high-field regime $(B \gg B_0^*)$, the dwell time is short and 
inter-arc scattering is ineffective, so the tunnel magnetoconductance 
coincides with the clean-interface result and is essentially insensitive 
to disorder. In contrast, in the low-field regime $(B \ll B_0^*)$, disorder 
can modify the slope of the linear magnetoconductance due to enhanced 
scattering and partial equilibration between transport channels.

Importantly, the universal conductance saturation at large magnetic 
fields remains robust against disorder, since it is ultimately 
determined by the number of available chiral edge modes in the layered 
Chern insulator, which are topologically protected and insensitive to 
weak interface scattering.

Experimentally, the predicted conductance  plateau could be probed 
in heterostructures combining WSMs with layered quantum anomalous Hall 
systems or magnetic topological insulator multilayers \cite{Burkov_Balents_2011,
Wang_2015, Bosnar2023}, or in engineered WSM–WSM interfaces 
\cite{Abdulla_pairwise_2024, Abdulla2026} tuned to the appropriate regime.

Overall, our results highlight how the interplay between chiral 
Landau levels and momentum-resolved Chern edge modes at topological 
interfaces can produce robust and universal transport behavior
controlled primarily by topology rather than microscopic 
band-structure details.


\begin{acknowledgements}
NB acknowledges the  International Centre for Theoretical Sciences (ICTS), where 
a  significant part of this work was done, for their hospitality, funding, and 
kind support toward academic collaboration. 
\end{acknowledgements}


\bibliography{main}

\end{document}